# The Potential of Geminate Pairs in Lead Halide Perovskite revealed via Time-resolved Photoluminescence

*Hannes Hempel[1]\*, Martin Stolterfoht[2], Orestis Karalis[1], Thomas Unold[1]*

[1] Department of Structure and Dynamics of Energy Materials, Helmholtz-Zentrum Berlin für Materialien und Energie, Hahn-Meitner-Platz 1, 14109 Berlin, Germany.

[2] Electronic Engineering Department, The Chinese University of Hong Kong, Hong Kong SAR, China.

\*Corresponding author: Email: hannes.hempel@helmholtz-berlin.de

***Photoluminescence (PL) under continuous illumination is commonly employed to assess voltage losses in solar energy conversion materials. However, the early temporal evolution of these losses remains poorly understood. Therefore, we extend the methodology to time-resolved PL, introducing the concepts of geminate PL, doping PL, and sibling PL to quantify the transient chemical potential of photogenerated electron-hole pairs and key optoelectronic properties. Analyzing the initial PL amplitudes reveals "hot charge carrier" separation for around 100 nm and is likely limited by the grain size of the triple cation perovskite. The following PL decay is caused by the diffusive separation of non-excitonic geminate pairs and time-resolves a fundamental yet often overlooked energy loss by increasing entropy. For triple-cation halide perovskite, we measure a "geminate correlation energy" of up to ~90 meV, persisting for ~ten nanoseconds. This energy is unutilized in standard solar cells and is considered lost in the Shockley-Queisser model. Therefore, this geminate energy could substantially enhance the device's efficiency, particularly under maximum power point and low-illumination conditions.***

The Shockley-Queisser (SQ) limit is widely regarded as the upper bound for the power conversion efficiency of conventional single-junction solar cells.(*1*) This limit accounts for various losses, including insufficient light absorption, charge carrier thermalization, radiative recombination, and increasing entropy. While the former losses are studied in great detail for all solar cell technologies, the entropic losses are far less studied and understood.(*2–4*) Entropy is associated with disorder and the uncertainty of where to find photogenerated charge carriers. To understand the entropic losses, we consider that photon absorption first generates a free geminate electron-hole pair, which is not bound as an exciton. Geminate pairs have two often overlooked and entangled consequences: Firstly, the order associated with the spatial correlation (overlap) of the geminate electron and hole pair contains additional free energy, which is not considered in the SQ-limit and may be harnessed. Secondly, the correlation also increases the photoluminescence amplitude, which allows measuring this extra energy by the generalized Planck's law.(*5*) The separation of geminate carriers reduces the free energy by increasing entropy and quenches the photoluminescence. This process is partially responsible for the difference between the bandgap energy of the semiconductor and the usable energy of a photogenerated electron-hole pair, i.e. the chemical potential, or the quasi-Fermi-level splitting, which limits the actual voltage of the solar cell. This loss will be monitored here by time-resolved photoluminescence (trPL).

Until today, the trPL was primarily studied to determine charge carrier lifetimes and recombination rate coefficients in perovskites.(*6–10*) In rare cases, implied current-voltage curves and power conversion efficiencies,(*10*) charge carrier mobilities and diffusion coefficients,(*11*) doping concentrations,(*12*) intrinsic carrier concentrations,(*6*) and external radiative coefficients(*13*) were determined. Recently, Krückemeier et al. measured the evolution of the quasi-Fermi-level splitting by trPL and compared it to external transient voltages of solar cells (*14*). Also, the trPL due to geminate recombination was barely studied. Only Augulis et al. demonstrated that the initial PL is governed by geminate recombination in lead halide perovskites at ps-timescales and low injection levels.(*15*) Although this work significantly changes the interpretation of trPL, the perovskite community mostly ignored the impact of geminate PL in subsequent works.

Our work integrates these aspects and may change the perception of trPL by showing its real potential: Firstly, it determines almost all key optoelectronic properties, far beyond lifetimes, even accessing hot carrier transport lengths. Secondly, it directly time-resolves energy losses by recording the transient chemical potential after pulsed photogeneration. Thirdly, it reveals the additional energy due to geminate correlation and surface photoexcitation.

We find that the entropic energy loss occurs partially during carrier thermalization due to hot carrier transport over a distance of ~100 nm, which seems limited to the grain size. The remaining extra geminate energy of up to 100 meV lives extraordinarily long (~10 ns), allowing for much slower (and possibly easier) energy conversion than in hot carrier approaches. Such geminate devices would harness the additional free energy from the correlation of the geminate pairs, allowing them to overcome the SQ-limit and would increase the power conversion efficiency of real devices, especially at low light intensities. Our findings are important for solar cells operated under low-intensity conditions with widespread application for internet-of-things applications. This new concept and the precise knowledge of the energy losses during the power conversion process are crucial for lead halide perovskites solar cells as they rapidly approach their SQ-limit and currently have (behind GaAs) the lowest relative voltage deficit compared to their SQ-limit.(*1*, *16*, *17*) Beyond perovskite solar cells, these considerations are likely applicable to organic solar cells and many other thin-film technologies.

**The Initial Amplitude of Photoluminescence Transients**

For the present study, we choose a common triple cation lead halide perovskite $Cs_{0.05}(FA_{0.83}MA_{0.17})_{0.95}Pb(I_{0.83}Br_{0.17})_3$ spin-coated on glass and encapsulated with a glass cover sheet, because it is possibly the best-studied halide perovskite composition, relatively stable, and has been widely used in very similar form in perovskite-silicon tandem solar cells.(*18*, *19*)

We measured trPL transients (Figure 1a) on the perovskite film for an increasing pulsed laser intensity, which photogenerated sheet carrier concentrations $\Delta n_s$ from $2 \times 10^7$ to $2 \times 10^{13}$ cm$^{-2}$ or mean carrier concentrations $\langle \Delta n \rangle = \Delta n_s / \langle d \rangle$ from ~$10^{12}$ to $10^{18}$ cm$^{-3}$. Here, we start with the relatively simple modeling of the initial amplitude of the trPL transients (Figure 2b) by Equation (1), which determines the dominating PL types, the external radiative coefficient, and doping concentration or geminate distance. Equation (1) combines the known equations for geminate,(*15*) sibling,(*13*) and doping PL.(*12*) We define these PL types in the method section.

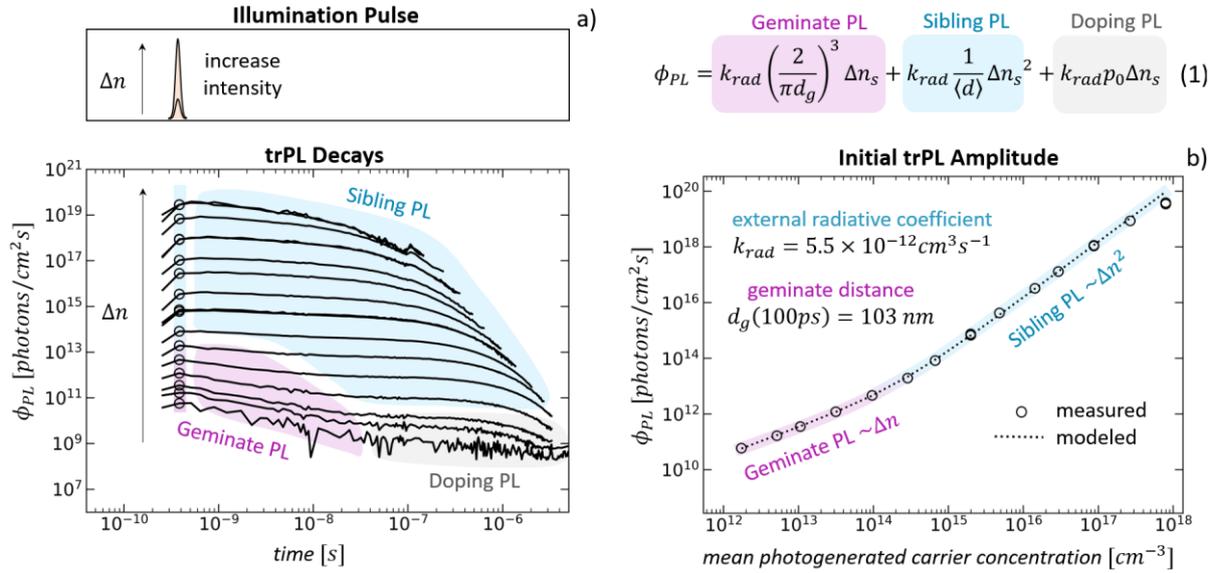

*Figure 1: Initial PL amplitude and QFLS. a) trPL transients measured for increasing pulsed photoexcitation from $6 \times 10^7$ to $3 \times 10^{13}$ photons/cm²/pulse on an encapsulated $Cs_{0.05}(FA_{0.83}MA_{0.17})Pb(I_{0.83}Br_{0.17})_3$ thin film. The dominant PL type is indicated by color. b) The initial amplitude of the trPL transient reveals Sibling PL ($\sim \Delta n^2$) at large photogenerated carrier concentration and geminate PL ($\sim \Delta n$) at low carrier concentration. Modeling with Equation 1 determines the external radiative coefficient $k_{rad}$ and the geminate distance $d_g$ at the time resolution of the trPL (~100ps).*

At strong illumination, corresponding to high injection conditions, the measured initial trPL amplitude increases quadratically with the mean photogenerated carrier concentration, which implies dominating sibling PL as the other types increase linearly. Fitting by Equation 1 determines an external radiative coefficient of $5.5 \times 10^{-12} cm^3 s^{-1}$ for a mean distribution depth $\langle d \rangle \geq 2/\alpha$ given initially by the absorption coefficient $\alpha$ and broadened by hot transport to 260 nm (Figures S1d and S4b). Simbula et al. determined similar radiative coefficients by this approach (Supporting Figure S2a) but neither accounted for hot transport nor resolved the geminate PL at low intensities.(*13*) Previously, one order of magnitude larger (Supporting Figure S2b) external radiative coefficients were often determined from the decay (not the amplitude) of trPL or other transient techniques.(*6, 10, 20*) However, this approach does not distinguish between radiative and non-radiative recombination coefficients. Therefore, it tends to overestimate as will be detailed later (Figure 3c), an ongoing discussion in the literature.

At lower carrier concentrations, the initial trPL amplitude increases linearly with photogenerated carrier concentration, which agrees with doping PL(*12*) and geminate PL.(*15, 21*) A distinction between them is possible by the kinetics: At the carrier concentration where the linear power-dependence of the PL amplitude occurs, a fast initial decay component appears in the measured trPL in Figure 2a. This is the expected behavior for free geminate PL and will be modeled later by the diffusive separations of free geminate pairs. Augulis et al. showed that this fast initial decay disagrees with charge trapping into defects.(*15*) In contrast to geminate or sibling PL, doping PL is unaffected by the diffusion of charge carriers (Figure S3c) and dominates in the gray area in Figure 2b, where the PL decays rather slow. Hence, the initial trPL amplitude at low intensity is attributed to geminate PL and is modeled with an initial geminate carrier distance $d_g$ of 103 nm. Since the Doped PL must have a lower amplitude than the measured PL the doping concentration must be below 10$^{14}$ cm$^{-3}$ and will be determined precisely later in this work.(*12*)

In combination with the measured absorptance $a$ (Figure S1b) and the black body spectrum $\varphi_{BB}(E)$ (Figure S1c) additional fundamental semiconductor properties can be determined. Without illumination, the sample emits only thermal radiation $\phi_{thermal} = \int \varphi_{BB}(E)a(E)\,dE$ of 0.037 photons/m²/s. In combination with the trPL-derived radiative coefficient and the film thickness $l$ of 410 nm, an intrinsic carrier concentration $n_i = \sqrt{\phi_{thermal}/k_{rad}l}$ of 1.2x10$^5$ cm$^{-3}$ is obtained. It translates for the bandgap of 1.65 eV to a joint effective density of states $N_j = n_i exp(E_G/2k_BT)$ of 8.7x10$^{18}$ cm$^{-3}$.(*6*)

**Diffusive Separation of Lone Geminate Pairs**

Now we model the trPL decays and start with an ultra-low photoexcitation at which geminate PL dominates. At this excitation level, only ~1 electron-hole pair / µm²/ pulse is generated, as illustrated in the secondary electron microscopy (SEM) image of the perovskite film in Figure 3a. With an grain size of ~120 nm, in only one out of 500 grains an electron-hole pair is photo-generated per pulse. Therefore, the geminate electron and hole will initially dominantly recombine radiatively with each other and will not interact with other (non-geminate) charge carriers. Hence, the initial trPL in Figure 2b is dominated by geminate PL. Its decay is caused by the increasing geminate electron-hole distance and is modeled by diffusion with a sum diffusion coefficient of $D_\Sigma = 0.0017\ cm^2s^{-1}$. Previously, geminate PL was observed in halide perovskites only up to 80 ps, likely due to limitations of the used PL-upconversion setup.(*15*) In contrast, we show here, that geminate PL dominates up to ~10 ns. For later times, using geminate PL, and also sibling PL, is insufficient to model the measured trPL. Additional doping PL from a doping concentration of $p_0 = 4.2\ x\ 10^{12}\ cm^{-3}$ is needed, when the external radiative coefficient is fixed to the value from fitting the initial trPL amplitude. Both, the determined doping concentration and the sum mobility $\mu_\Sigma = D_\Sigma e/(k_BT)$ of 0.64 cm²V$^{-1}$s$^{-1}$ implied by the sum diffusion coefficient agree with our previous Hall measurements ($p_{0,Hall} = 1.2\ x\ 10^{12}cm^{-3}$, $\mu_{Hall} = 1.7\ cm^2V^{-1}s^{-1}$) on equivalent samples.(*12*) The trPL decay at even longer times is attributed to recombination with an effective lifetime of 1.2 µs.

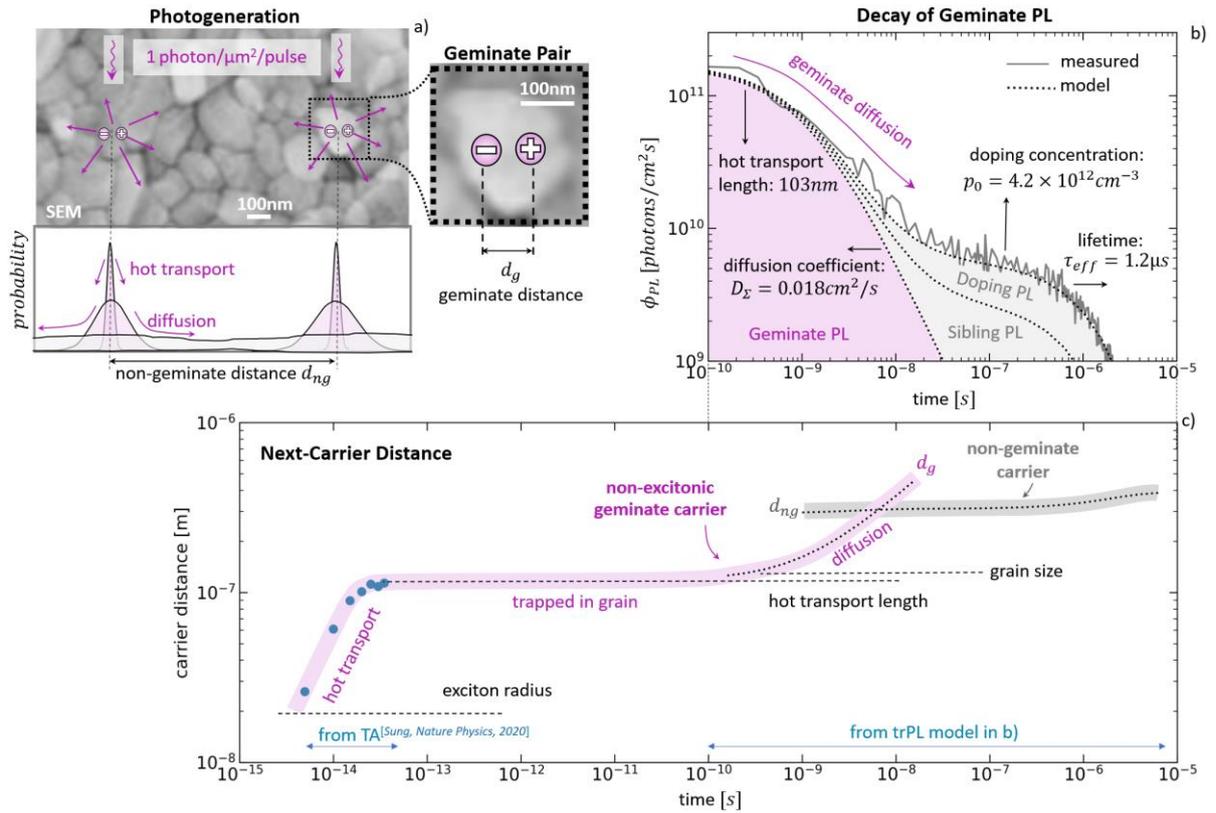

*Figure 2: Diffusion of non-excitonic geminate electron-hole pairs. a) SEM image of the polycrystalline perovskite film superimposed with an illustration of the photogeneration of individual geminate electron-hole pairs at an ultra-low photon flux of only ~1 photon/µm²/pulse. b) Zoomed SEM illustrates the distance between the geminate electron and hole at the time resolution of the trPL measurement of ~100ps, which seems limited to the grain size. c) The measured trPL transient is modeled by Equation (2). The initial geminate PL decays (purple area) by separating the geminate carrier pairs by diffusion with a diffusion coefficient of 0.018 cm²V⁻¹. The PL tail is dominated by Doped PL from a doping concentration $p_0$, which decays with an effective charge carrier lifetime $\tau_{eff}$ due to recombination. d) The mean distance to the next charge carrier from modeled trPL and the transient absorption (TA) data of Sung et al.. (22) Until ~10 ns the non-excitonic geminate partner is the closest next carrier and the geminated PL is dominant.*

A trPL decay by charge transport can be understood as an increase in the mean distance between a photogenerated carrier and the next charge carrier, which decreases the chance of its radiative recombination. The distance to the geminate partner $d_g$ and to the next non-geminate majority carrier $d_{ng}$ are illustrated in Figure 2a and shown in Figure 2c. They are determined from the trPL model (Figure 2b) and complemented by transient absorption (TA) data published by Sung et al..(22) After 10 fs, the TA-derived next carrier distance is larger than the reported exciton radius of 3-10 nm.(23) Therefore, and because of the exciton binding energy of ~10-30 meV comparable to the thermal energy, we observe the transport of non-excitonic free geminate carriers. The geminate distance increases to 50-200 nm (depending on the sample in Sung et al.) in less than ~100 fs, attributed to the ballistic transport of hot charge carriers. This agrees with the geminate distance of 103 nm determined here from the initial trPL amplitude in Figure 1b and identifies it as the hot carrier transport length. A comparison with typical grain sizes (Figure 2a) indicates, that hot carriers possibly transport ballistically until they scatter at the grain boundaries, which cools them to lattice temperature. Then, the mean next-carrier distance is in the order of the grain size and remains constant from ~100 fs to ~1 ns. After 1ns, the geminate carriers diffuse across different grains. Hence, the trPL-derived diffusion coefficient (see Figure 2b) should be understood as an inter-grain value. In contrast, the intra-grain transport can be significantly faster, as shown by the intra-grain mobilities of ~30 cm²V⁻¹s⁻¹ measured previously by terahertz spectroscopy on equivalent samples or long-range photoconductivity in single crystals.(*10, 24*) At ~10 ns, the geminate distance becomes longer than the (2D) mean distance to the next non-geminate carrier $d_{ng} = 0.5 p_s^{1/2}$.(*25, 26*) This distance of ~300 nm is comparable to the thickness of the perovskite film and the non-geminate carriers start behaving like a 2D distribution, where $p_s$ is the sheet carrier concentration of the majority carriers.

The diffusion of non-geminate photogenerated carriers from the initial absorption profile to a homogeneous distribution causes for sibling PL a trPL decay by a factor of $\approx 2/\alpha_{eff} l$, where $\alpha_{eff}$ is the effective absorption coefficient and $l$ is the film thickness. In contrast, Doping PL would be unaffected by this factor as detailed in the SI. Modeling the initial Sibling trPL decay measured at moderate illumination (Figure S3a) requires broadening the initial carrier profile by hot transport, which agrees with the results of the geminate PL. The underlying carrier diffusion is modeled by the continuity Equation S24 and determines a diffusion coefficient (0.017 $cm^2 s^{-1}$) in line with the result of the geminate PL and with the results from analyzing transient PL spectral shifts by Cho et al. ($0.01 - 0.02\ cm^2 s^{-1}$).(*27*)

In general, diffusion of initially confined distributions causes PL decays, which approach $\sim t^{-b/2}$, where the exponent $b$ depends on the PL type (Figure S3c). In our case, the geminate PL decays with $b = 3$, due to the diffusion in all three dimensions, whereas the sibling PL decays with $b = 1$ due to the lateral homogeneous photoexcitation and the diffusion only into the dimension of light propagation. Neglecting PL reabsorption, doping PL does not decay by diffusion ($b = 0$).

**Transient Chemical Potential and the Geminate Energy**

Previously, implied voltages (and chemical potentials of photogenerated electron-hole pairs $\mu_{eh} = eV$) were often determined by the generalized Planck's law from the steady-state PL calibrated to absolute units of [photons/cm²/s].(*5*) Equation 2 is the integral version, which was alleraady used in slightly different form by Shockley and Queisser. It is tempting to interpret calibrated trPL with this equation as a measurement of the transient chemical potential of the electron-hole pairs, as previously done by Krückemeier et al..(*14*)

$$\mu_{eh}(t) \approx k_B T \ln\left(\frac{\phi_{PL}(t)}{\phi_{thermal}}\right) \tag{2}$$

Inserting Equations 1 for PL and S19 for thermal emission yields the chemical potential generalized for geminate correlation (Equation (3)), where the three summands in the square bracket correspond to geminate, doping, and sibling PL, respectively.

$$\mu_{eh}(x,t) = k_B T \ln\left(\left[\left(\frac{2}{\pi d_g(t)}\right)^3 + p_0 + \Delta n(x,t)\right]\frac{\Delta n(x,t)}{n_i^2}\right) = qV \qquad (3)$$

By averaging over inhomogeneous non-geminate carrier distributions $\Delta n(x,t)$ we can derive the mean transient chemical potential of photogenerated electron-hole pairs $\langle\mu_{eh}\rangle(t)$:

$$\langle\mu_{eh}\rangle(t) = \frac{\int \Delta n(x,t)\, \mu_{eh}(x,t)\, dV}{\int \Delta n(x,t)\, dV} \qquad (4)$$

This derivation is rather heuristic as Planck's generalized law was derived for homogeneous non-geminate carrier distributions and its validity is not shown for geminate carriers or inhomogeneous distributions. Hence, we derive in the Supporting Note S8 the chemical potential of a geminate pair from its entropy, and Equation 3 seems valid.

The transient chemical potential of photogenerated electron-hole pairs (Figure 3a) is estimated by generalized Planck's law directly from the trPL of Figure 2b and modeled by Equations 3 and 4. It exhibits an initial decay by ~100 meV, corresponding to the diffusion-related decay of the trPL, and a decay on timescales beyond 100 ns, corresponding to the recombination-related decay of the trPL.

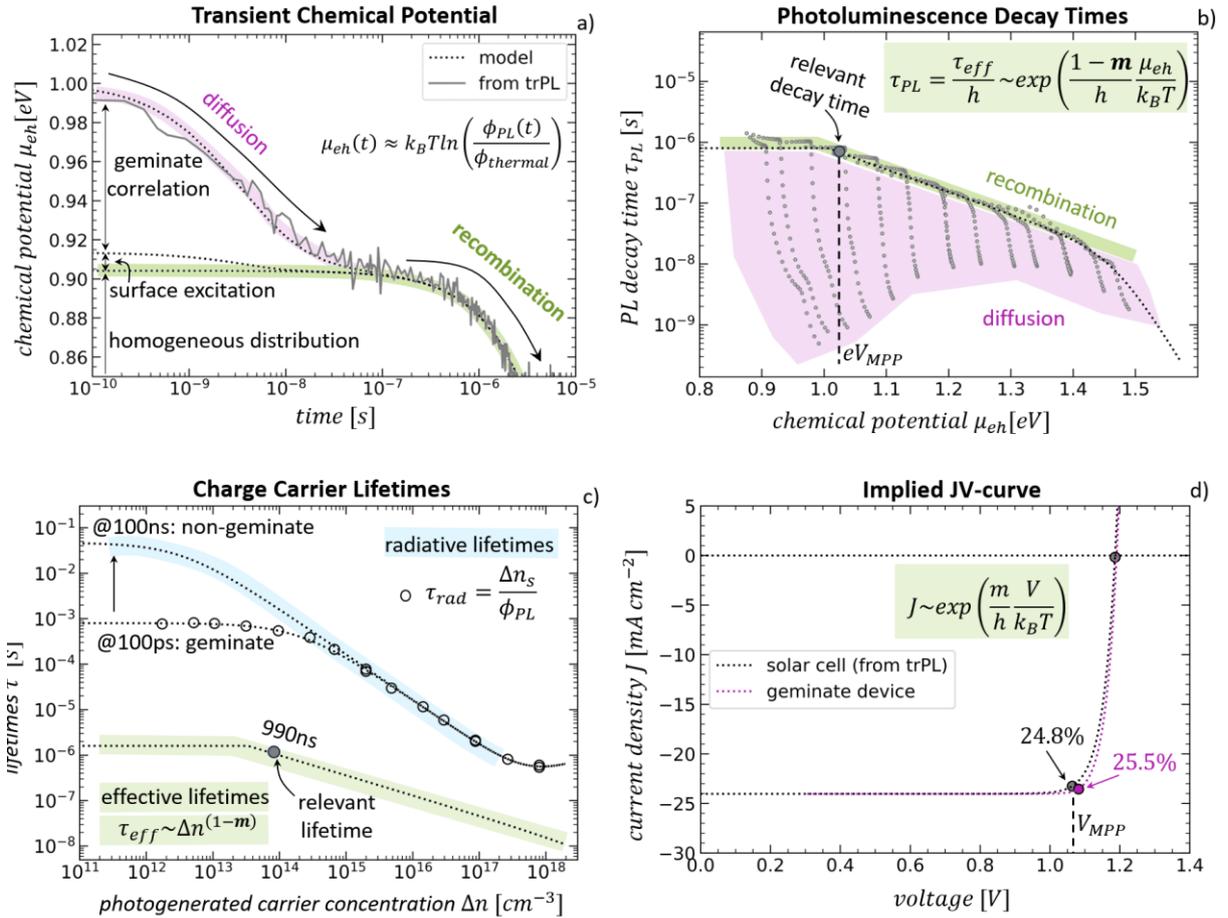

*Figure 3: The transient Chemical Potential of Geminate Pairs a) The chemical potential is modeled from the evolution of the carrier distributions and approximated by the so-called "generalized Planck's law" from the measured trPL. The geminate correlation and surface excitation imply additional free energy, which is lost during the diffusion. The homogeneous chemical potential decays by recombination. b) The differential PL decay times are derived from the trPL decay kinetics and are shown as a function of the chemical potentials derived from the trPL amplitudes. The voltage at maximum power point ($V_{MPP}$) of the implied JV-curve determines the decay time relevant for PV application. c) The current-voltage curve of a solar cell and geminate device (which utilizes geminate energy) from detailed balance analysis yields the power conversion efficiencies implied by the measured trPL.*

We model the transient chemical potential for 3 cases (dotted lines in Figure 3a) based on the evolution of the carrier distributions and geminate distances known from modeling the trPL. The first case starts with a homogeneous distribution of photoexcited carriers and neglects geminate correlation, resulting in a homogenous initial chemical potential of 0.905 V decaying only by charge carrier recombination.

The second case includes an inhomogeneous carrier generation profile neglecting geminate correlation and results in a slightly (~10 meV) larger initial chemical potential, which decays within ~10 ns to the values of the homogeneous distribution by diffusion.

The third case considers the generation profile and includes geminate correlation by a finite $d_g$. Equation 3 describes the trPL-derived chemical potential in Figure 3a well. The geminate correlation adds another ~90 meV to the initial chemical potential, which we define as geminate energy $E_{geminate} = \mu_{eh}(d_g) - \mu_{eh}(d_g = \infty)$ and decays within ~ 20 ns by diffusion.

A decay of the chemical potential by diffusion is caused by increasing entropy, due to the increase in possible microstates (positions) in which the carriers are. The spread out of the geminate pair and the non-geminate redistribution throughout the film are fundamental losses in free energy, in the chemical potential of the electron-hole pairs, in their Fermi-level splitting, and finally in the implied voltage that the probed material could supply in a solar cell with ideal charge transport layers. To our knowledge, the loss of geminate correlation has been overlooked in the literature on energy conversion devices. (*2–4*) Remarkably, this loss occurs on the nanosecond time scale, many orders of magnitude slower than the energy loss during charge carrier thermalization. Whereas a hot carrier solar cell must convert energy within picoseconds before thermalization,(*28*) the geminate energy could be converted much slower.

As the generalized Planck's law is roughly valid, the amplitudes of all trPL transients shown in Figure 2a are transformed into chemical potentials, and their decays into PL decay times (Equation S45), as introduced by the Kirchartz group and shown in Figure 3b.(*9*) The decay times from different injection levels that fall on the same (green) line are caused by charge carrier recombination. The decay times that do not fall on this line are caused by diffusion, and the loss of geminate correlation or non-geminate redistribution causes the decrease of the chemical potentials in this area.

The recombination-related PL decay times translate to effective charge carrier lifetimes $\tau_{eff} = \tau_{PL} h$ with a factor $h = 2$ for sibling PL or $h = 1$ for doping PL (Figure 3c). The external radiative lifetimes are obtained from the measured PL amplitudes $\tau_{rad} = \Delta n_s / \phi_{PL}$. They follow $\tau_{rad}^{-1} = k_{rad} \Delta n$ at high injection and approach $\tau_{rad}^{-1} = k_{rad} p_0$ at low injection and after the separation of geminated pairs. The effective charge carrier lifetimes are much shorter and therefore not limited by radiative recombination. The effective lifetimes decrease with increasing chemical potential (larger carrier concentration) and are described by $\tau_{eff} \sim \Delta n^{-(m-1)}$ with an unconventional non-radiative recombination process of order *m* of 1.44. Conventional recombination processes such as trap-assisted, radiative, or Augere recombination have integer values of *m*. Hence, unconventional non-radiative higher-order recombination limits the probed perovskites. This recombination models all measured trPL transients with one set of parameters (Table S2, Figure S4a).

Regardless of the origin of this recombination, a detailed balance analysis, described in detail in our previous work by Hempel et al.,(*10*) yields the implied current-voltage curve (Figure 3d) that the materials can supply in a solar cell under solar illumination. We also measured the photoluminescence quantum yield $Y_{PL} = \tau_{eff}/\tau_{rad}$ under continuous illumination, which agrees perfectly with the trPL-determined effective and radiative lifetimes (Figure S4c). This agreement proves that trPL correctly predicts the charge carrier dynamics under continuous illumination. The trPL-implied power conversion efficiency for the halide perovskite sample investigated is 24.8%. The voltage $V_{mpp}$ at the maximum power point (MPP) and the corresponding chemical potential of 1.063 eV determine the PL decay time of 470 ns and corresponding charge carrier lifetime of 940 ns relevant for photovoltaic application. Hence, such lifetimes at the MPP should be stated to judge photovoltaic materials. Even perovskite samples with lifetimes of 270 μs reported at low injection levels have lifetimes of ~ 1 μs at 1-sun illumination equivalent MPP and will have an implied power conversion efficiency similar to our sample.(*9*) The photogenerated carrier concentration of $8 \times 10^{13}$ cm$^{-3}$ at the MPP is much larger than the doping concentration. Therefore, doping does not support the voltage at the implied MPP.

While it is not obvious how to use the geminate energy in a conventional solar cell because it requires the separation of electrons and holes, the geminate energy may be used by processes where an electron and a hole are converted at the same place e.g. by photochemistry, or exciton splitting, or light emission. It must be a process, that converts the energy of the geminate pair before it diffuses apart. Using the chemical potential after thermalization (with the measured geminate distance of 102 nm) in the detailed balance analysis (see supporting note S10) and assuming its complete conversion to an external voltage, improves the fill factor of the implied current-voltage curve and increases the efficiency to 25.5 %. This boost in chemical potential and device efficiency is more pronounced at lower light intensities (lower carrier concentrations) and smaller geminate distances (Figure S4d). For example, the geminate energy of ~90 meV determined in Figure 3a was measured at a photogenerated carrier concentration of only ~$10^{12}$ cm$^{-3}$, which corresponds to ~0.003 suns-equivalent illumination and directly increases the implied $V_{oc}$ of the geminate device by ~90 mV and the efficiency from 18.5 % to 22.0 % (Figure S5a).

The ballistic transport of hot carriers may be beneficial for carrier extraction, which however is usually not limiting perovskites solar cells. On the downside, it increases the geminate distance to 102 nm, thereby reducing the chemical potential. If the geminate distance could be reduced to 10 nm, e.g. by faster carrier cooling and reduced ballistic transport, the radiative efficiency limit of the geminate device would increase from 29.3 % for the conventional solar cell to 31.9 % and hence surpass the Shockley-Queisser limit (Figure S5c).

**Conclusion**

This work shows that distinguishing between geminate PL, sibling PL, and doping PL is essential for interpreting and modeling trPL. The associated differences and accessible optoelectronic properties are summarized in Table 1. Halide perovskites are particularly suited for observing geminate PL of free carriers, because of their relatively low exciton binding energy, low charge carrier mobility, long lifetime, low doping, and strong PL even under weak illumination. Dominating geminate PL is indicated by a linear dependence of the initial trPL amplitude on illumination intensity and a fast initial decay from diffusive geminate separation.

*Table 1: Difference between PL types and accessible material properties*

|  | Sibling PL | Doping PL | Geminate PL |
|---|---|---|---|
| Dominates for | strong illumination | strong doping | slow diffusion slow recombination |
| Excitation intensity dependence | quadratic | linear | linear |
| Decay by diffusion | slow ~$t^{-0.5}$ | no decay | fast ~$t^{-1.5}$ |
| Decay time by recombination | half the lifetime | lifetime | lifetime |
| Unique accessible properties | radiative coefficient $k_{rad}$ diffusion coefficient $D$ | doping concentration $p_0$ | hot transport length $d_{hot}$ diffusion coefficient $D$ |

other accessible properties: $\tau_{eff}, \tau_{rad}, \mu_\Sigma, n_i, N_j, J(V), \eta$

We have shown that the initial trPL amplitude allows the measurement of the external radiative coefficient $k_{rad}$ and radiative lifetimes $\tau_{rad}$. Additionally, the intrinsic carrier concentration $n_i$ and the joint effective density of states $N_{eff}$ are obtained by combining trPL and absorptance data. The initial trPL decay up to ~10 ns is dominated by charge transport while the later component is dominated by recombination. For a trPL time resolution of ~100 ps, a mean distance between the geminate electron and hole of ~100 nm is determined from the trPL amplitude and supported by TA results. This distance arises from a separation of carriers during thermalization due to ballistic carrier transport and is identified as the hot carrier transport length, possibly limited by the comparable grain size. After 1 ns carriers diffuse across grains with a diffusion coefficient of 0.018 cm²s⁻¹ (mobility of 0.8 cm²V⁻¹s⁻¹). The trPL-derived diffusion coefficient and doping concentration of 3x10¹²cm⁻³ are in agreement with Hall measurements, which highlights the potential of trPL beyond characterizing the effective charge carrier lifetimes $\tau_{eff}$.

This study demonstrates that trPL can be interpreted (with some restrictions) as a time-resolved measurement of the chemical potential of photogenerated electron and hole pairs, which limits the external voltage of solar cells. It identifies energy losses by recombination and charge transport, and allows estimation of the implied JV curve under solar illumination. This curve is consistent with PLQY measurements and suggests a potential power conversion efficiency of 24.8% for the probed neat $Cs_{0.05}(FA_{0.83}MA_{0.17})_{0.95}Pb(I_{0.83}Br_{0.17})_3$ perovskite.

In addition, the analysis of trPL reveals that the spatial correlation of the electron-hole pair generated by the same photon implies an energy not utilized in traditional solar cells. This "geminate energy", for which we measure up to 90 meV, is lost during the diffusive separation of the geminate pair and has been overlooked as a fundamental energy loss process in solar cells so far. Preventing this loss would significantly increase the power conversion efficiency at low light intensities and even allow slightly exceeding the Shockley-Queisser limit. Remarkably, this loss occurs on very slow time scales (nanoseconds) compared to the thermalization losses (femtoseconds), potentially opening up new possibilities for harnessing this energy.

**Method Section:**

*Modeling trPL*

During the trPL measurement, the sample (lead-halide perovskite thin film encapsulated with glass) is illuminated with a series of short laser pulses, which photogenerate electron-hole pairs as illustrated in Figure 1. Partially, these charge carriers recombine radiatively, emitting photoluminescence (PL). The emitted photon flux $\phi'_{PL}$ in units of photons/s can be described by Equation (5), which contains the external radiative coefficient $k_{rad}$ and the product of the electron concentration $n(r,t)$ and the hole concentration $p(r,t)$. This product reflects that an electron and a hole must be present at the same position simultaneously to recombine radiatively. Therefore, PL decays are caused either by recombination, which reduces the overall number of carriers, or by transport, which reduces their spatial overlap when they diffuse away from each other or are separated by electric fields or contact layers.

$$\phi'_{PL}(t) \approx k_{rad} \int_0^V n(r,t)p(r,t)dV \quad \left[\frac{photons}{s}\right] \quad (5)$$

Therefore, describing trPL requires modeling the evolution of the electron and hole distributions, for which we use the continuity equation S24, which connects photogeneration, recombination, diffusion, and drift of charge carrier distributions as detailed in the SI. The carrier distributions can be described in two spatial frames: non-geminate and geminate. The halide perovskite community, except of Augulis et al., has mostly neglected the geminate frame so far.(*15*)

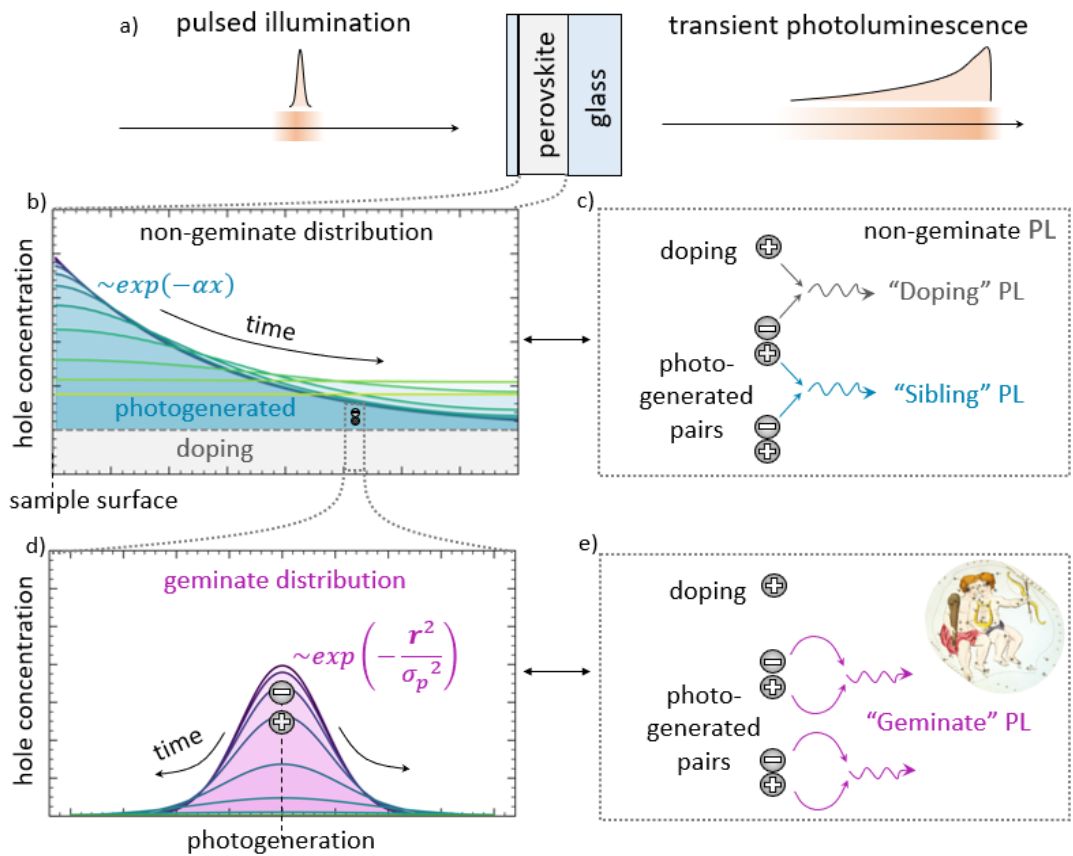

*Figure 4: Theory of Geminate and Non-geminate Photoluminescence.* a) Illumination of a semiconductor with a laser pulse results in photogeneration of charge carriers and subsequent radiative recombination, which emits photoluminescence transients, which are described in Equation (3) by the electron and the hole distributions $n$ and $p$. b) The non-geminate representation of the charge carrier distributions is fixed to the sample surface/ geometry. It captures e.g. the photogeneration profile by an exponential decrease towards the depth of the sample and a homogeneous doping. These non-geminate distributions describe the non-geminated PL illustrated in c), which can be distinguished in Doped PL (a photogenerated carrier radiatively recombines with a doping carrier) and sibling PL (both recombining carriers were photogenerated by different photons). d) The non-geminate representation of the carrier distributions is fixed to the position of the photogeneration of each carrier. It captures the spatial correlation of geminate electron-hole pairs (generated by the same photon at the same position) with a Gaussian distribution and describes the geminated PL illustrated in e).

Usually, the carrier distribution is regarded in the spatial frame of the sample (Figure 4b) with the origin of the x-axis at the sample surface. This frame captures the "non-geminate" distribution of the photogenerated carrier concentrations $\Delta p = \Delta n = \Delta n_0 \, exp(-\alpha x)$ right after photogeneration, which decreases with the absorption coefficient $\alpha$ into the depth of the sample. Over time, these carriers will diffuse into the depth of the sample and recombine. In addition, a doping concentration $p_0$ (holes in this case) may be present, which are the carriers already present in the dark equilibrium. Hence, non-geminate photogenerated charge carriers can either radiatively recombine with another photogenerated carrier that was generated by another photon, which we call here "sibling PL", or with a doped charge carrier, which we call here "doping PL". We introduce this distinction because they behave differently as detailed in the main section. In this work, we only detect the PL where the non-geminate carrier distributions are laterally homogeneous, which simplifies the modeling to one spatial dimension and yields the non-geminate PL (in the gray part of Equation (6)) in units of photons/cm²/s.

$$\phi_{PL}(t) \approx \underbrace{k_{rad} \int_0^l n(x,t)p(x,t)dx}_{\substack{\text{numerical modeling} \\ \text{of } n(x,t)p(x,t)}} + \underbrace{k_{rad} \left(\frac{2}{\pi d_g(t)}\right)^3 \Delta n_s(t)}_{\substack{\text{analytic solution for} \\ \text{Gaussian distribution}}} \quad \left[\frac{photons}{cm^2 s}\right] \tag{6}$$

(non-geminate PL) (geminate PL)

However, the absorption of a photon generates an electron and a hole at the same position called a geminate pair. The term "geminate" originates from the astrological sign Gemini, which shows twins. The geminate charge carriers are in that sense twins generated by the same photon. If the geminate carrier pair recombines radiatively with each other, it emits geminate PL. To describe the spatial correlation of a non-excitonic geminate carrier pair, we regard a second spatial frame centered (Figure 4d) at the position of the photogeneration. We describe the geminate distributions with 3-dimensional Gaussians.

$$d_g(t) = \sqrt{\frac{2^3}{\pi}}\sigma_g = \sqrt{\frac{16}{\pi}D_\Sigma t + {d_o}^2} \tag{7}$$

The geminate PL of a single geminate pair is derived by solving analytically Equation (3) in the SI and depends on the mean distance between the geminate electron and the geminate hole $d_g$. This geminate distance increases in Equation (5) by diffusion with the sum $D_\Sigma = D_n + D_p$ of the electron diffusion coefficient $D_n$ and the hole diffusion coefficient $D_p$, and an initial distance $d_o$. The geminate

PL of geminate pairs with a sheet concentration $\Delta n_s = \int \Delta n \, dx$ is given by the purple part Equation (6).

This equation allows modeling PL by drift, diffusion, and recombination of non-geminate charge carrier distributions, complemented by geminate contribution.

*trPL Setup:*

The sample is photoexcited by a pulsed femtosecond laser with a wavelength of 515 nm, a pulse length of 400 fs, a repetition rate of 125 kHz, and a beam diameter of 3 mm. The light intensity is adjusted by reflective neutral density filters. The incident light intensity [W/cm$^2$] is measured by a power meter to which a pinhole with a diameter of 0.5 mm is attached. The laser light is suppressed by a 550 nm long pass and a 50 nm bandpass centered at 750 nm. The trPL is measured by a single photon counting system from PicoQuant with a Geiger-mode avalanche photodiode with a detector size of 50 µm and measured with a time resolution of 100 ps.

*Sample information*:

*Perovskite solutions:* The triple cation perovskite solution was prepared by mixing two 1.2 M FAPbI$_3$ and MAPbBr$_3$ perovskite solutions in DMF:DMSO (4:1 volume ratio, v:v) in a ratio of 83:17 which we call "MAFA" solution. The 1.2 M FAPbI$_3$ solution was thereby prepared by dissolving FAI (722 mg) and PbI2 (2130 mg) in 2.8 mL DMF and 0.7 mL DMSO which contains a 10 molar% excess of PbI2. The 1.2 M MAPbBr$_3$ solution was made by dissolving MABr (470 mg) and PbBr$_2$ (1696 mg) in 2.8 mL DMF and 0.7 mL DMSO which contains a 10 molar% excess of PbBr2. Lastly, 40 µL of a 1.5 CsI solution in DMSO (389 mg CsI in 1 mL DMSO) was mixed with 960 µL of the MAFA solution resulting in a nominal perovskite stoichiometry of $Cs_{0.05}(FA_{0.83}MA_{0.17})_{0.95}Pb(I_{0.83}Br_{0.17})_3$.

*Perovskite film fabrication:* All triple cation perovskite films were deposited on fused silica by spin-coating at 4500 rpm for 35 s. After 10 s of the start of the spinning process, the spinning substrate was washed with 300 µL ethylacetate for approximately 1 s (the anti-solvent was placed in the center of the film). We note, that by the end of the spinning process the perovskite film turned dark brown. The perovskite film was then annealed at 100 °C for 1 h on a preheated hotplate where the film turned slightly darker.

**Acknowledgments:**

*Author contributions:*

Conceptualization: HH. Methodology: HH. Investigation: HH, OC. Visualization: HH. Funding acquisition: TU. Project administration: HH. Supervision: HH, TU. Writing – original draft: HH. Writing – review & editing: HH, MS, TU.

*Competing interests:*

The authors declare that they have no competing interests.

*Data and materials availability:*

All data are available in the main text or the supplementary materials. The Python code for modeling and analysis can be requested from the corresponding author.

# Supporting Information

# *The Potential of Geminate Pairs in Lead Halide Perovskite revealed via Time-resolved Photoluminescence*

*Hannes Hempel[1]\*, Martin Stolterfoht[2], Orestis Karalis[1], Thomas Unold[1]*

[1] Department of Structure and Dynamics of Energy Materials, Helmholtz-Zentrum Berlin für Materialien und Energie, Hahn-Meitner-Platz 1, 14109 Berlin, Germany.

[2] Electronic Engineering Department, The Chinese University of Hong Kong, Hong Kong SAR, China.

\*Corresponding author: Email: hannes.hempel@helmholtz-berlin.de

## S1: Theory of non-geminate PL

The non-geminate PL $\phi_{ng}$ is calculated by Equation 5 of the main manuscript, which is here simplified to Equation S1 by assuming that the distribution of doped carriers is homogeneous and that the photogenerated electron and holes follow the same distribution, which is in general true right after photogeneration. $l$ is the thickness of the film.

$$\phi_{ng} = k_{rad} \int_0^l n(x)p(x)dx \approx k_{rad} \int_0^l \Delta n(x)[\Delta n(x) + p_0]dx \tag{S1}$$

Illumination of the sample photogenerates electrons and holes with a distribution $\Delta n(x)$, where $\Phi_{laser}$ is the flux of the pulsed laser in [photons /cm$^2$/pulse], $R$ is the reflectance of 23 % (Figure S1b), and $\alpha$ is the absorption coefficient of 1.2x10$^5$cm$^{-1}$ at the wavelength of 515 nm of the exciting pulsed laser:

$$\Delta n(x) = (1-R)\Phi_{laser}\alpha\, exp(-\alpha x) \tag{S2}$$

It induces a sheet carrier concentration $\Delta n_s$ given by Equation S3. The approximation in the following equation S3-S6 is for the absorption profile in Equation S2 and $\alpha l \gg 1$:

$$\Delta n_s = \int_0^l \Delta n(x)dx \approx (1-R)\Phi_{laser}\,[1-exp(-\alpha l)] \approx (1-R)\Phi_{laser} \tag{S3}$$

The mean photogenerated majority carrier concentration that a photogenerated minority carriers encounter is:

$$\langle \Delta n \rangle = \frac{\int_0^l \Delta n(x)\Delta p(x)dx}{\int_0^l \Delta n(x)dx} = \frac{(1-R)^2 \Phi_{laser}^2 \frac{\alpha}{2}[1-exp(-2\alpha l)]}{\Delta n_s} \approx \Delta n_s \frac{\alpha}{2} \quad \text{(S4)}$$

This allows defining the mean distribution depth $\langle d \rangle$ of the photogenerated charge carriers, which increases for a broader distribution (lower mean carrier concentration at constant sheet carrier concentration):

$$\langle d \rangle = \frac{\Delta n_s}{\langle \Delta n \rangle} \approx \frac{2}{\alpha} \quad \text{(S5)}$$

Combing Equations S1-S5 yields Equation S6 for the non-geminate PL:

$$\phi_{ng} = k_{rad}\frac{1}{\langle d \rangle}\Delta n_s^2 + k_{rad}p_0\Delta n_s \approx k_{rad}\frac{\alpha}{2}\Delta n_s^2 + k_{rad}p_0\Delta n_s \quad \text{(S6)}$$

The Sibling PL (first term) decays with increasing distribution depth (diffusion into a broader distribution) or decreasing overall sheet carrier concentration by recombination. The Doping PL (second term) decays only by recombination, not diffusion.

Alternatively to the absorption profile, one may assume that the carriers are distributed homogeneously over the film thickness $l$, which leads with $\langle d \rangle = l$ to a photoluminescence flux $\phi_{homo}$ of:

$$\phi_{homo} = k_{rad}l^{-1}\Delta n_s^2 + k_{rad}p_0\Delta n_s^2 \quad \text{(S7)}$$

Hence, redistribution from the absorption profile to the homogeneous distribution leads for Sibling PL and for $\alpha l \gg 1$ to a decay in Sibling PL by:

$$\frac{\phi_{homo}}{\phi_\alpha} = \frac{2}{l\alpha} \quad \text{(S8)}$$

For the initial trPL amplitude (at the time-resolution of ~100ps) the carrier distribution given by the absorption profile is broadened by hot carrier transport, which is detailed in Figure S3b and leads to a mean distribution depth between $2/\alpha \leq \langle d \rangle \leq l$.

## S2: Theory of Geminate PL

The distribution of the photogenerated geminate electron $n_g(t,r)$ and of the geminate hole $p_g(t,r)$ are described in the spatial frame centered at the position of the photogeneration of the geminate pair. Here we assume that they can be described by 3 dimensional normal distributions with individual variances $\sigma_n$ and $\sigma_p$, which can be identified with the root mean square displacements from the center of their common photogeneration.

$$n_g(x,y,z) = \frac{1}{(2\pi)^{1.5}\sigma_n(t)^3} exp\left[-\frac{x^2+y^2+z^2}{2\sigma_n^2}\right] \tag{S9}$$

$$p_g(x,y,z) = \frac{1}{(2\pi)^{1.5}\sigma_p^3} exp\left[-\frac{x^2+y^2+z^2}{2\sigma_p^2}\right] \tag{S10}$$

The probability distribution $\rho_g(\Delta x, \Delta y, \Delta z)$ of having the geminate partner position $\Delta x, \Delta y, \Delta z$ is a convolution of the electron and hole distribution. A convolution of normal distributions yields also a normal distribution with an increased variance of $\sigma_g^2 = \sigma_n^2 + \sigma_p^2$. The probability of having the geminate partner at a distance $d(\Delta x, \Delta y, \Delta z) = \sqrt{\Delta x^2 + \Delta y^2 + \Delta z^2}$ from to geminate partner is:

$$\rho_g = \frac{1}{(2\pi)^{1.5}\sigma_g^3} exp\left[-\frac{d^2}{2\sigma_g^2}\right] \quad \text{with} \quad \sigma_g^2 = \sigma_n^2 + \sigma_p^2 \tag{S11}$$

The mean distance between the geminate carriers is the local probability $\rho_g(d)$ times the local distance $d$ averaged over all 3 spatial dimensions (volume) in which the geminate partner may be located:

$$\langle d_g \rangle = \int_0^\infty d\, \rho_g\, dV = \sqrt{\frac{2^3}{\pi}}\sigma_g = \left(\frac{2^4}{\pi} D_\Sigma t\right)^{0.5} \tag{S12}$$

Diffusion increases the variance of a Gaussian distribution by $\sigma = (2Dt)^{0.5}$. The mean geminate distance increase with the sum diffusion coefficient $D_\Sigma$ given by the sum of the electron diffusion coefficient $D_n$ and the hole diffusion coefficient $D_p$. They can be translate by the Einstein relation to the corresponding mobilities.

$$D_\Sigma = D_n + D_p = \frac{k_B T}{e}\mu_\Sigma \tag{S13}$$

The total geminate PL $\phi_g$ is the product of this PL of a single geminate pair and the sheet concentration of photogenerated pairs:

$$\phi_g = \Delta n_s k_{rad} \int n_g(x,y,z) p_g(x,y,z)\, dV \tag{S14}$$

Inserting the Gaussian distributions (Equations S9 and S10) into Equation S14 yields:

$$\phi_g(t) = \Delta n_s(t) k_{rad} \frac{1}{(2\pi)^3 \sigma_n^3 \sigma_p^3} \int_{-\infty}^{\infty} exp\left[-\frac{(x^2+y^2+z^2)}{2\frac{\sigma_n^2 \sigma_p^2}{\sigma_p^2+\sigma_n^2}}\right] dV \tag{S15}$$

Solving and inserting the mean distance between the geminate pair yields:

$$\phi_g(t) = \Delta n_s(t) k_{rad} \frac{1}{(2\pi)^{1.5} \sigma_g(t)^3} = \Delta n_s(t) k_{rad} \left(\frac{2}{\pi \langle d_g \rangle(t)}\right)^3 \tag{S16}$$

Augulis et al. state that "the effective delocalization ($\sigma$) of the carrier wave functions can be obtained from the ratio of quadratic and linear components" of the PL intensity, which correspond to the Sibling PL and the Geminate PL. The ratio of Sibling PL the Geminate PL by Equation S16 and S6 yields Equation S16b, which is similar to Equation 1 in Augulis paper apart of a factor of 2, which may be connected to their definition of the "effective delocalization".

$$\sigma_g = \frac{1}{(2\pi)^{1/2}} \left(\frac{\phi_{ng}}{\Delta n \phi_g}\right)^{1/3} \quad for\ \Delta n \gg p_0 \tag{S16b}$$

The total PL is the sum of geminate PL (Equation S16) and non-geminate PL (Equation S6), which leads to Equation 1 of the main manuscript.

### S3 Measured Photon Absorption and Emission

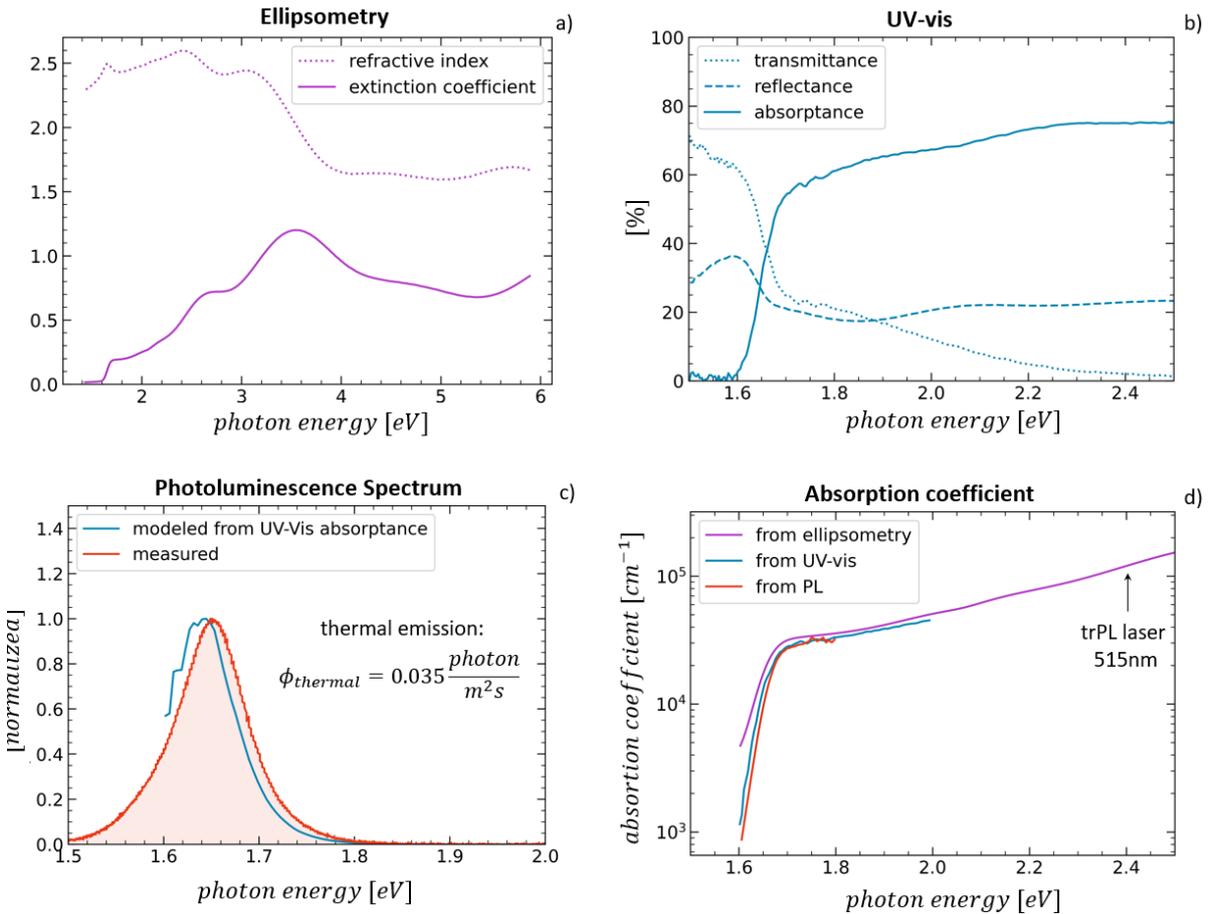

*Figure S1:* a) Refractive index and extinction coefficient from fitting Psi and Delta values of ellipsometry data. b) transmittance and reflectance measured in an Ulbricht sphere. c) Measured PL and thermal photon emission derived from the absorptance. The emitted thermal photon flux by Kirchhof's law is 0.035 photons/m$^2$/s. d) Absorption coefficients derived from a), b), and c). The absorption coefficient is obtained from ellipsometry, transmittance and reflectance measurements (UV-vis), and photoluminescence (PL).

We determined the absorption coefficient in three different ways which closely agree on each other (Figure S1d): by ellipsometry, by transmittance and reflectance measured in the UV-vis, and by PL.

The absorption coefficient is calculated from the reflectance $R$, the transmittance $T$, and the film thickness $l$.

$$\alpha = -\frac{1}{l} ln\left(\frac{T}{1-R}\right) \tag{S17}$$

The absorption coefficient is also calculated from the measured PL spectrum $\varphi_{PL}(E)$ based on Planck's generalized law, where $c$ is the speed of light, and $h$ is Plank's constant.

$$\alpha = -\frac{1}{l} ln\left(\varphi_{PL}(E) \frac{h^3 c^2}{2\pi E^2} exp\left(\frac{E - \mu_{eh}}{k_B T}\right)\right) \tag{S18}$$

## S4 The external radiative coefficients and lifetimes:

The external radiative coefficient $k_{rad}$ differs from the internal radiative coefficient $k'_{rad}$ by e.g. PL reabsorption and PL outcoupling.(*6*) For example, a smoother surface or a thicker sample would decrease the external radiative coefficient, whereas the internal coefficient would remain constant. However, the external radiative coefficient is preferred here as it avoids rather complex optical modeling (with many free parameters) such as the matrix method and ray tracing.

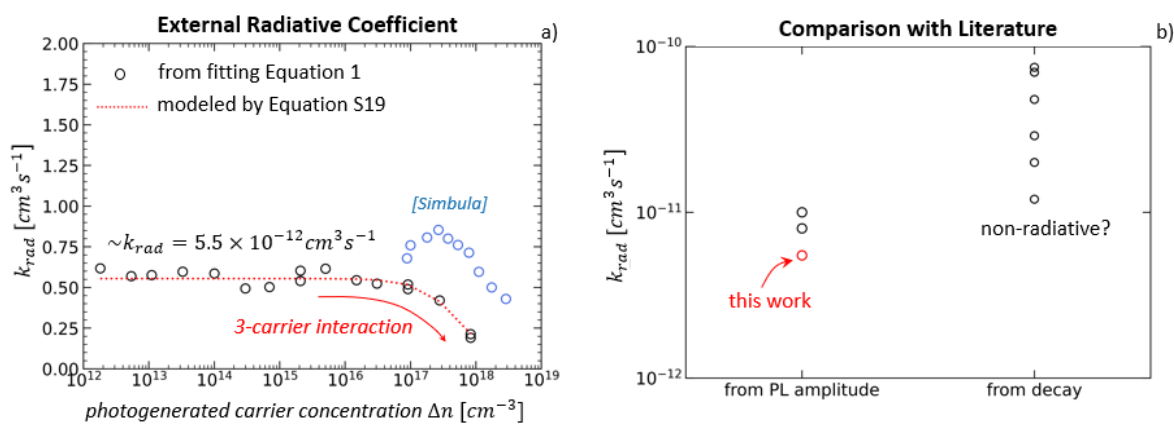

*Figure S2: External radiative coefficient. a) From fitting the initial trPL amplitude (Figure 1b) and combined with literature data from Simbula et al. (13) b) Comparison with literature data. Analyzing PL amplitude yields lower values than analyzing decay kinetics.*

We fitted the initial trPL amplitude (Figure 1b main manuscript) by Equation 1 of the main manuscript with the external radiative coefficient as a free parameter for each photogenerated carrier concentration, a fixed distribution depth of 264 nm, and a fixed geminate distance of 104 nm. The retrieved radiative coefficients (Figure S2a) are constant at low photogenerated carrier concentrations and start decreasing above ~ $10^{17} cm^{-3}$. This decrease was observed by Simbula et al. before and can be explained by the filling of band states, exciton screening, Auger recombination, or other processes where three carriers are involved. (*13*) As a phenomenological description of this injection dependence we used Equation S19. Below a critical carrier concentration $b_1$ of 7x10$^{17}$cm$^{-3}$ the external radiative coefficient is constant with a value $k_{rad,0}$ of $5.5 \times 10^{-12} cm^3 s^{-1}$ and for larger carrier concentrations it decreases with an exponent $b_2$ of 1.5.

$$k_{rad} = \frac{k_{rad,0}}{1 + \left(\frac{\Delta n}{b_1}\right)^{b_2}} \tag{S19}$$

The distribution depth used here is broadened by hot carrier transport, as will be detailed in Figure S3b. Neglecting such additional broadening of the carrier distribution would result in a distribution depth of only 167 nm (Equation S5) given by the absorption coefficient and the radiative coefficient obtained from fitting the initial trPL amplitudes would increase to a low-injection value of $8.7 \times 10^{-12} cm^3 s^{-1}$ in excellent agreement with the values from Simbular et al., which indeed neglected hot carrier broadening.

Alternatively, the external radiative lifetime can be used instead of the external radiative coefficient. It is given by Equation S20 and is shown in Figure 3c of the main manuscript.

$$\tau_{rad}^{-1} = \frac{\phi_{PL}}{\Delta n_s} = k_{rad} \left[\frac{\Delta n_s}{\langle d \rangle} + p_0 + \left(\frac{2}{\pi d_g}\right)^3\right] \tag{S20}$$

The intrinsic carrier concentration can be determined from the thermal emission $\phi_{thermal}$, the external radiative coefficient $k_{rad}$, and the thickness of the film $l$.

$$n_i = \sqrt{\frac{\phi_{thermal}}{k_{rad} l}} \tag{S21}$$

*Table S1: Recombination Coefficients in Literature:*

| Method | $k_2$ [cm$^3$s$^{-1}$] | $k_3$ [cm$^6$s$^{-1}$] | reference |
| --- | --- | --- | --- |
| TA transient | 2.9E-11 | 1E-30 | (*20*) |
| PL transient | 4.8E-11 | 8.8E-29 | (*6*) |
| TRMC transient | 7E-11 | | (*10*) |
| THz transient | 2E-11 | | (*29*) |
| TA transient | 7-8.1 E-11 | | (*30*) |

| | | | | |
|---|---|---|---|---|
| PLQY | 0.9-1.6 E-11 | | | (30) |

## S5 non-geminate next carrier distance

The mean distance to the nearest neighbor for a homogenous (infinite) distribution with carrier concentration p in the 3-dimensional space is: (26)

$$\langle d_{ng} \rangle = 0.554 p^{1/3} \tag{S22}$$

However, for 2-dimensional space, it is given by the sheet carrier concentration of the majority carriers $p_s$: (25)

$$\langle d_{ng} \rangle = 0.5 p_s^{1/2} \tag{S23}$$

Halide perovskite thin films are usually viewed as 3D materials as there is no quantum confinement for a thickness of ~500 nm. However, the carrier distribution becomes in another sense 2D-like, when the nearest neighbor distance is longer than the film thickness at very low carrier concentrations. Then the delocalization of the carrier is confined in one dimension by the layer thickness and the nearest neighbor distance should be calculated by Equation S23.

## S6: Modeling non-geminate charge carrier dynamics

The continuity equation models the evolution of the carrier distributions $n$:

$$\frac{d}{dt} n = \frac{d}{dx} I_{diff} - R + G \tag{S24}$$

The recombination $R$ is given by the effective charge carrier lifetime $\tau_{eff}$ of the photogenerated minority carrier concentration $\Delta n$:

$$R = \frac{1}{\tau_{eff}} \Delta n \tag{S25}$$

The diffusion current $I_{diff}$ is given by the diffusion coefficient and the gradient of the carrier concentration.

$$I_{diff} = -D \frac{d}{dx} n \tag{S26}$$

The effective charge carrier lifetime is described here phenomenologically by a recombination coefficient $k_m$ and a recombination order m. The effective lifetime is limited to a maximum value $\tau_{max}$, which implies a fixed lifetime $\tau_{max}$ at low carrier concentration.

$$\frac{1}{\tau_{eff}} = k_m \Delta n^{m-1} \quad for \ \tau_{eff} < \tau_{max} \tag{S27}$$

## S7 Impact of Charge Transport on Individual PL Types

Now we turn to a trPL transient recorded at a moderate photoexcitation of 7.4 x $10^{10}$ photons/cm². In Figure 1b it was shown that at this intensity, Sibling PL dominates, for which carrier redistribution must lead to a decay of the trPL as e.g. simulated by Fai et. al. (*31*).

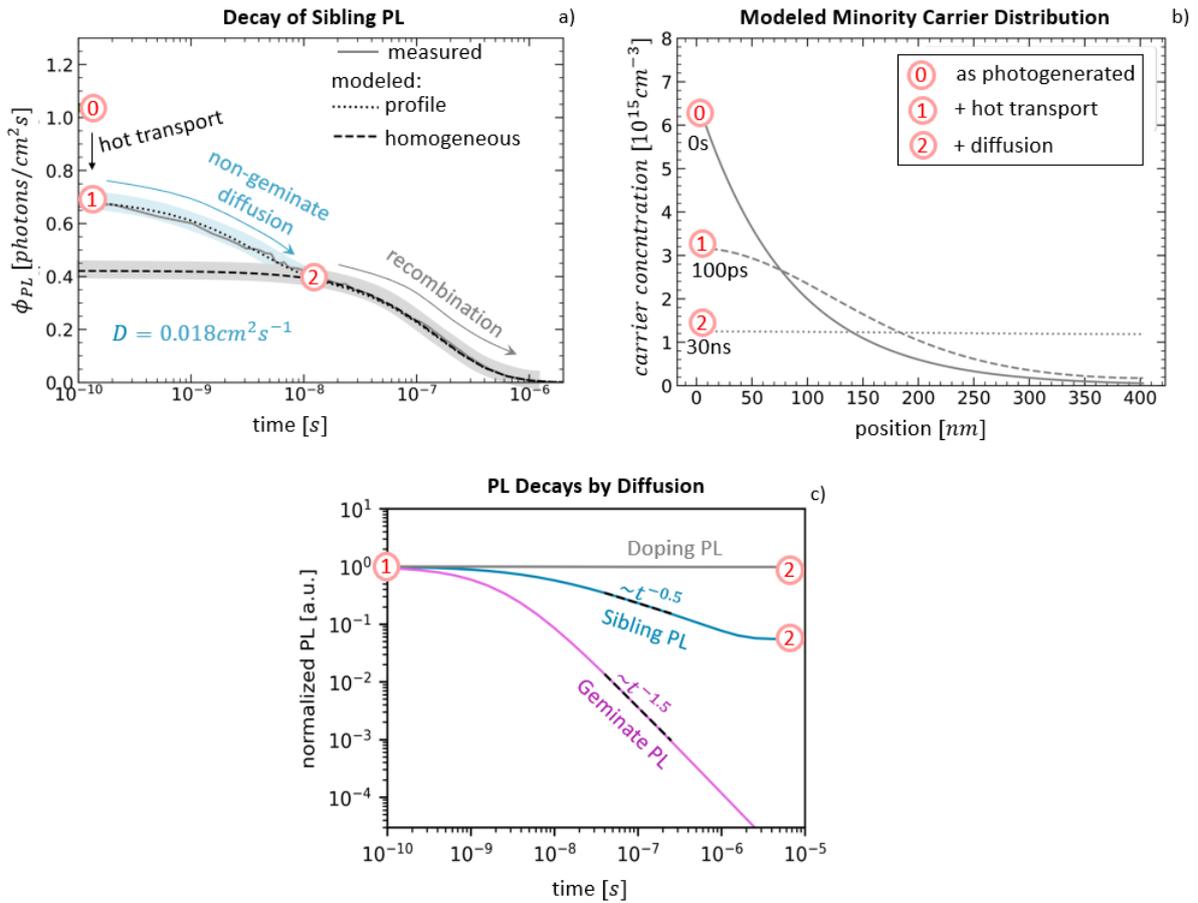

*Figure S3: Transport-induced PL decays a) trPL measured at photoexcitation of 7.4 x $10^{10}$ photons/cm² where Sibling Pl dominates. Modeled with an initially homogeneous distribution and the absorption profile broadened by hot carrier transport. b) carrier distributions of (0) the absorption profile, (1) the broadened absorption profile imitating hot carrier transport, and (2) the homogeneous distribution after 30 ns of diffusion. c) Modelling the different PL types for a 5 µm thick film with surface excitation, diffusion, and no recombination.*

The measured trPL transient exhibits two decay components in Figure S4a. The first component can be attributed to diffusion and the second to recombination. The trPL modeled from a homogenous photoexcitation describes only the later decay by charge carrier recombination with an effective lifetime of 400 ns. The initial decay is modeled with a diffusion coefficient of 0.017 cm²s⁻¹. This diffusion coefficient controls how fast the initial trPL decays and is in close agreement with the diffusion coefficient obtained from modeling the Geminate PL at much lower carrier concentrations (Figure 2b of the main script). Figure S3b shows as "0" the photogenerated carrier distribution from the illumination by the laser, which follows the Lambert-Beer-like absorption profile with a mean carrier distribution depth of 167 nm ($\langle d \rangle = 2/\alpha$) given by the absorption coefficient $\alpha$ and as "2" the final homogenous distribution after diffusion throughout the layer thickness $l$ with a distribution depth of 410 nm ($\langle d \rangle = l$). The decay of the trPL amplitude is given by the ratio of the distribution depths, which is a decay to 41 % ($l\alpha/2$) of the initial trPL amplitude. However, the observed decay to 65 % is smaller suggesting a larger "initial" distribution depth of 265 nm. This larger distribution depth and broader carrier distribution can be attributed to hot carrier transport, which is not resolved by trPL. To implement this broadening, we allow the carrier distribution to diffuse for broadening $\sigma = 2Dt$ of 107 nm, before the actual modeling starts.

While the diffusion coefficients obtained from Geminate and Sibling PL are similar, their PL decays are different. To compare the impact of diffusion on the different PL types (geminate PL, doping PL, and sibling PL), we modeled in Figure S3c the trPL of a perovskite film with the same properties but a thickness of 5 µm and without recombination. Geminate PL is quenched by diffusion into all three spatial dimensions and approaches a decay of $\sim t^{-1.5}$. Sibbing PL is in our case quenched only by diffusion into the depth of the sample as the sample is laterally homogeneously illuminated and the decay approaches $\sim t^{-0.5}$. In general, the decay approaches $\sim t^{-d/2}$, where $d$ is the number of dimensions into which the carriers diffuse. In contrast, Doping PL does not decay by diffusion, if the doping carriers are homogeneously distributed and if reabsorption of PL is neglectable. Note that for doping PL, the mean Fermi level of the minority carriers will still decay by diffusion although the trPL stays constant. Hence, in this case, the generalized Planks law seems to fail to predict correctly the transient chemical potential of the electron and hole pairs from the Photoluminescence.

## S8 Theory of the chemical potential

For non-geminate carriers, the chemical potential at position *x* is given by the Boltzmann constant $k_B$, the temperature *T* and the intrinsic carrier concentration $n_i$. (*2*)

$$\mu_{eh}(x,t) = k_B T \, ln\left(\frac{n(x,t)p(x,t)}{n_i^2}\right) \tag{S28}$$

If the carriers are distributed inhomogeneous, their chemical potential will depend on the position. To calculate the mean chemical potential of all photogenerated carriers $\langle\mu_{eh}\rangle$, we assume that the distributions of photogenerated electrons $\Delta n(x)$ and holes $\Delta p(x)$ equal each other (same diffusion coefficient, no drift).

$$\langle\mu_{eh}\rangle(t) = \frac{\int \Delta n(x,t)\, \mu_{eh}(x,t)\, dV}{\int \Delta n(x,t)\, dV} \tag{S29}$$

It leads to:

$$\langle \mu_{eh} \rangle(t) = \frac{k_B T}{\Delta n_s} \int \Delta n(x,t) \, ln\left(\frac{n(x,t)p(x,t)}{n_i^2}\right) dV \quad \text{(S30)}$$

In a more general case, the difference between the mean chemical potential of the photogenerated electrons and the mean chemical potential of the photogenerated holes would have to be calculated, which leads to:

$$\langle \mu_{eh} \rangle(t) = k_B T \left[\frac{\int \Delta n(x,t) \, ln\left(\frac{n(x,t)}{n_i}\right) dV}{\int \Delta n(x,t) \, dV} + \frac{\int \Delta p(x) \, ln\left(\frac{p(x,t)}{n_i}\right) dV}{\int \Delta p(x,t) \, dV}\right] \quad \text{(S31)}$$

However, this non-geminate chemical potential assumes that the electron and the hole are not correlated and therefore underestimates the chemical potential of a geminate electron-hole pair. It may be tempting to use the geminate distributions for electrons and holes in Equation S9 and S10 to regard geminate correlation. However, this approach would incorrectly assume that electrons and holes are always photogenerated at the same position and overestimate the chemical potential. Instead, we suggest, using the non-geminate distribution for the one carrier species (the majority carriers) and the geminate distribution for the other carrier species (the minority carriers). This approach reflects the entropy of the geminate charge carrier pair and the knowledge about the probability of where (in which microstate) they are: The probability of finding the eighter the carriers is proportional to the non-geminate distribution and the probability of finding its geminate partner in a certain distance is proportional to the geminate distribution.

The chemical potential of an electron-hole pair is given by the change of the free energy $dG$ by a change in the particle number $dN$. The decay of the chemical potential is caused by an increase of the entropy $S(t)$.

$$\mu_{eh}(t) = \frac{dG}{dN} = E_G + E_{kin} + pV - T\frac{dS(t)}{dN} = E_G + 5k_B T - T\frac{dS(t)}{dN} \quad \text{(S32)}$$

The entropy of a system is :

$$S = -k_B \sum_i w_i \, ln(w_i) = -k_B N \int w(\Omega) \, ln(w(\Omega)) \, d\Omega \quad \text{(S33)}$$

We are transitioning from the summation over a discreet set of a number of $i$ states with probability $w_i$ to the integration over a continuum of states with an effective density $N$, which are occupied with a probability $w$ that depends on a variable $\Omega$. The number of states in an interval $d\Omega$ is $Nd\Omega$.

For non-geminate homogeneous distributions of electron and holes, the entropy was derived by Sackur and Tetrode (*32*), (*33*), (*2*):

$$\frac{dS}{dN} = k_B \left(5 + ln\left(\frac{N_c N_v}{np}\right)\right) \quad \text{(S34)}$$

Here, we would like to generalize for inhomogeneous distributions and geminate correlation. However, a rigid derivation is beyond our capabilities and we will proceed with a rather heuristic argumentation: To reduce complexity, we consider a single photogenerated electron-hole pair. In a specific microstate, the electron is at position $(x, y, z)$ with probability $w_n$ and a density of (conduction band) states $N_c$, and the hole is at position $(x + \Delta x, y + \Delta y, z + \Delta z)$ with probability $w_p$ and a density of (valence band) states $N_v$. The probability $w = w_n w_p$ of this microstate and the effective density of states $N = N_v N_c$ lead with Equation S33 to an entropy of:

$$S = -k_B N_v N_c \int w_n w_p \, ln(w_n w_p) \, dV_n dV_p \tag{S35}$$

The probability $w_n$ that the electron is in a specific state at position $(x, y, z)$ is given by the (non-geminate) distribution of the electron $n(x)$ and the effective density of conduction band states $N_c$:

$$w_n(x, y, z) = \frac{n(x)}{N_c} \tag{S36}$$

The probability $w_p$ that the hole is in a specific state at position $(x + \Delta x, y + \Delta y, z + \Delta z)$ is given by the (geminate) distribution of the hole $p(x, \Delta x)$ and the effective density of valence band states $N_v$:

$$w_p(\Delta x, \Delta y, \Delta z) = \frac{p(\Delta x, \Delta y, \Delta z)}{N_v} = \frac{1}{(2\pi)^{1.5} \sigma_g^3 N_v} exp\left[-\frac{\Delta x^2 + \Delta y^2 + \Delta z^2}{2\sigma_g^2}\right] \tag{S37}$$

This leads to:

$$S = -k_B \int n(x) p(\Delta x, \Delta y, \Delta z) \, ln\left(\frac{n(x)}{N_c} \frac{p(\Delta x, \Delta y, \Delta z)}{N_v}\right) dV_n dV_p \tag{S38}$$

As we consider a single electron and a single hole:

$$\int n(x) \, dV_n = \int p(\Delta x, \Delta y, \Delta z) \, dV_p = 1 \tag{S39}$$

Note that $w_p$ only depends on the relative position (geminate frame), where as $w_n$ depends on the absolute position (non-geminate sample frame). Hence, both subsystems (for electron and for holes) are *independent (for the geminate correlation it does not matter where in the sample frame the geminate pair is)*, which leads to:

$$S = -k_B \int n(x) \, ln\left(\frac{n(x)}{N_c}\right) dV_n - k_B \int p(\Delta x, \Delta y, \Delta z) \, ln\left(\frac{p(\Delta x, \Delta y, \Delta z)}{N_v}\right) dV_p \tag{S40}$$

Inserting Equation S37 and solving yields:

$$S = S_n + S_p = -k_B \int n(x) \, ln\left(\frac{n(x)}{N_c}\right) dV_n - k_B \left(ln\left(\frac{1}{(2\pi)^{1.5} \sigma_g^3} \frac{1}{N_2}\right) - \frac{1}{2^3}\right) \tag{S41}$$

Now we assume that the variation in $n(x)$ is on a relatively large scale, much larger than the volume occupied by a single electron-hole pair. It allows treating the carrier concentration $n(x)$ as a constant during the integration of the volume $dV_n$ that contains a single electron. Inserting this entropy in Equation S32 and neglecting the $-2^{-3}$ yields:

$$\mu_{eh}(t,x) = k_B T \ln\left(\frac{1}{(2\pi)^{1.5}\sigma_g^3}\frac{n(x)}{n_i^2}\right) \quad (S42)$$

Without rigid proof, we combine Equation S42 and Equation S28 to derive a phenomenological description of the chemical potential:

$$\mu_{eh}(t,x) = k_B T \ln\left(\left[\frac{1}{(2\pi)^{1.5}\sigma_g(t)^3} + p(x,t)\right]\frac{n(x,t)}{n_i^2}\right) \quad (S43)$$

It includes both cases, the lone geminate pair and the non-geminate distribution, as limits. The mean chemical potential of the photogenerated electron-holes pairs obtained by the combination of Equations S43 and S29:

$$\langle\mu_{eh}\rangle(t) = \frac{k_B T}{\Delta n_s}\int \Delta n(x,t) \ln\left(\left[\frac{1}{(2\pi)^{1.5}\sigma_g(t)^3} + p_0 + \Delta n(x,t)\right]\frac{n(x,t)}{n_i^2}\right)dx \quad (S44)$$

## S9 Theory of Recombination:

An initial guess on the charge carrier lifetimes without modeling the trPL transients can be obtained by plotting the differential trPL decay times $\tau_{PL}$ versus the Fermi level splitting in Figure 3b. However, calculating such decay times with Equation S45 requires smoothing the trPL transient, which introduces some dependencies on the smoothing process.

$$\tau_{PL}(t) = -\phi_{PL}(t)\left(\frac{d}{dt}\phi_{PL}(t)\right)^{-1} = \left(\frac{d}{dt}\ln[\phi_{PL}(t)]\right)^{-1} \quad (S45)$$

Assuming homogeneous distribution (PL decay is not caused by diffusion), inserting generalized Plank's Law (Equation 2) and the thermal emission (Equation 21), and using the effective lifetime given by recombination for order m (Equation 27) yields:

$$\tau_{PL} = \frac{1}{h}\tau_{eff} \sim \frac{1}{h}\exp\left(\frac{[1-m]}{h}\frac{\mu_{eh}}{k_B T}\right) \quad (S46)$$

$h$ is a factor that depends on the PL type. $h = 2$ for Sibling PL ($\Delta n > p_0$) and $h = 1$ for Doping PL ($\Delta n < p_0$) and Geminate PL.

The effective charge carrier lifetimes and radiative lifetimes in Figure S4b and Figure 3c of the main manuscript are obtained from modeling all trPL transients in Figure S4a with the same parameter set of:

*Table S2: Parameters of the overall trPL model in Figure S4a*

|  | Symbol | Value |
|---|---|---|
| **Transport** | | |
| Diffusion coefficient | $D$ | $0.0017\ cm^2 s^{-1}$ |
| Initial non-geminate distribution depth | $\langle d \rangle$ | $260\ nm$ |
| Initial geminate distance | $d_g$ | $103\ nm$ |
| **External radiative coefficient** | | |
| Low-intensity value | $k_{rad,0}$ | $5.5\ x\ 10^{-12} cm^3 s^{-1}$ |
| Critical carrier concentration | $b_1$ | $5.5\ x\ 10^{17} cm^{-3}$ |
| Exponent | $b_2$ | $1.5$ |
| **Effective lifetime** | | |
| Maximum lifetime | $\tau_{max}$ | $1.6\ \mu s$ |
| Recombination order | $m$ | $1.44$ |
| Recombination coefficient | $k_m$ | $1.6\ x\ 10^{-3} cm^{3m} s^{-1}$ |
| **Other** | | |
| Doping concentration | $p_0$ | $4.2\ x\ 10^{12} cm^{-3}$ |
| Film thickness | $l$ | $410\ nm$ |

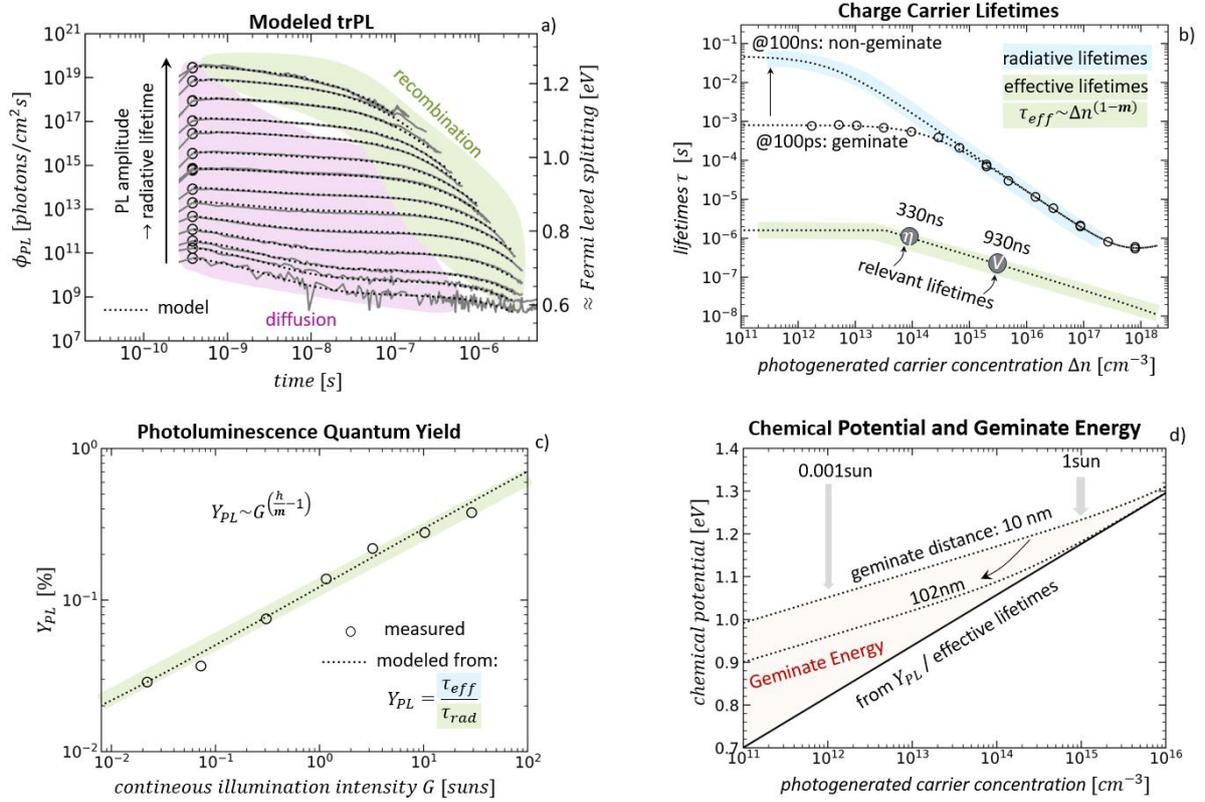

**Figure S4: Overall Model.** *a) trPL transients measured for increasing photogenerated carrier concentrations and modeled by the overall diffusion-recombination model with parameters from Table S2. b) The radiative and effective (non-radiative) charge carrier lifetimes used to model the trPL transients in a). c) The photoluminescence quantum yield measured under increasing continuous illumination is modeled very well by the radiation and effective lifetimes from b). d) The non-geminate chemical potential and additional geminate energy for different geminate distances.*

Under continuous illumination with a generation rate G a steady-state builds up with constant carrier concentrations.

$$\frac{d}{dt}\Delta n = 0 = G - \frac{1}{\tau_{eff}}\Delta n \quad (S47)$$

This steady-state carrier concentration without carrier extraction is:

$$\Delta n = G\tau_{eff} = G\frac{1}{k_m}\Delta n^{1-m} = \left(\frac{G}{k_m}\right)^{1/m} \quad (S48)$$

The continuous PL in this state is:

$$\phi_{PL}(t) \sim \Delta n(\Delta n + p_0) = \left(\frac{G}{k_m}\right)^{1/m}\left(\left(\frac{G}{k_m}\right)^{1/m} + p_0\right) \approx G^{h/m} \quad (S49)$$

The photoluminescence quantum yields $Y_{PL}$ is described by:

$$Y_{PL} = \frac{\tau_{rad}}{\tau_{eff}} = \frac{\phi_{PL}}{G} \sim G^{\left(\frac{h}{m}-1\right)} \quad (S50)$$

The photoluminescence quantum yields calculated from the trPL-derived radiative and effective lifetimes are in excellent agreement with the measured values in Figure S4c.

A conventional solar cell uses the chemical potential without geminate energy (Equation S28), whereas a geminate device contains geminate energy (Equation S43). Figure S4d shows that the additional geminate energy increases with decreasing photogenerated carrier concentration and decreasing geminate distance. For the measured geminate distance of ~100 nm, the geminate energy is relatively smaller for our sample under 1 sun illumination at implied MPP as the non-geminate potential is already large. However, for smaller geminate distances or at lower light intensities, the geminate energy is significantly larger and would also boost the power conversion efficiency of a geminate device.

## S10 Detailed Balance Analysis :

We use the detailed balance analysis for calculating the current - voltage curve that the probed material could yield in a power conversion device. The detailed balance states that all carriers that are generated by the generation $G$ and do not recombine by the recombination $R$ are extracted as a current density $J$:

$$J = -q(G - R) \quad (S51)$$

The Generation $G$ [carriers/cm²/s] is calculated here from the sun spectrum $\varphi_{sun}$, the absorptance $a$ and the concentration / dilution of the sun $c_{sun}$. It gives the short circuit current $J_{SC}$.

$$G = c_{sun}\int a\,\varphi_{sun}dE = J_{SC}/q \quad (S52)$$

The recombination $R$ [carriers/cm²/s] is given by the thermal photon emission $\Phi_{thermal}$, the photoluminescence quantum yield $Y_{PL}$ and the voltage $V$, implied by the Fermi-level splitting (chemical potential).

$$R = \frac{\Phi_{thermal}}{Y_{PL}(V)}\left[exp\left(\frac{qV}{k_BT}\right) - 1\right] \tag{S53}$$

The minus one originates from the absorption of thermal radiation and thermally generated carriers. This allows us to state the current-voltage curve:

$$J = -q\left(G - \frac{\Phi_{thermal}}{Y_{PL}(V)}\left[exp\left(\frac{qV}{k_BT}\right) - 1\right]\right) \tag{S54}$$

In the classical Shockley-Queisser (SQ) limit, the charge carriers recombine only via radiative recombination, and hence the photoluminescence quantum yield is unity. The absorptance is a step-like onset to unity at the bandgap and the sun spectrum is approximated by a 6000 K hot thermal emitter. The radiative limit uses the more realistic AM1.5G sun spectrum and the measured absorptance. Therefore it yields slightly different values.

However real materials, are usually dominated by non-radiative recombination, which is included in the photoluminescence quantum yield. Often and in particular of halide perovskites, it depends on the voltage. Hence in practice, it is hard to obtain the current-voltage characteristics from Equation S54.

For such cases and to include the geminate correlation, we used here the recombination as function of the photogenerated carrier concentration:

$$R = k_{rad}\Delta n(p_0 + \Delta n) + k_m \Delta n^m \tag{S55}$$

Assuming complete conversion of the chemical potential to an external voltage, this voltage is also stated as function of the photogenerated carrier concentration:

$$V \approx \frac{\mu_{eh}}{q} = \frac{k_BT}{q} ln\left(\left[\left(\frac{2}{\pi d_g}\right)^3 + p_0 + \Delta n\right]\frac{\Delta n}{n_i^2}\right) \tag{S56}$$

We calculate numerically the voltage-dependent recombination $R(V)$ by solving Equations S55 and S56 for a series of photogenerated carrier concentrations and use these values in Equation S51 to obtain the implied current-voltage curves, which regard the geminate correlation and voltage dependent photoluminescence quantum yields. At low light intensities of 0.003 suns (Figure S5a) the geminate device (22%) outperforms the conventional solar cell (18.5%) significantly due to a better Voc and fill factor. At solar illumination (Figure S5a) only the fill factor is improved for the measured geminate distance of 103 nm. However, a smaller geminate distance of 10 nm would significantly boost the geminate device efficiency. Even the radiative limit would be increased (Figure S5c) and surpassed the SQ-limit.

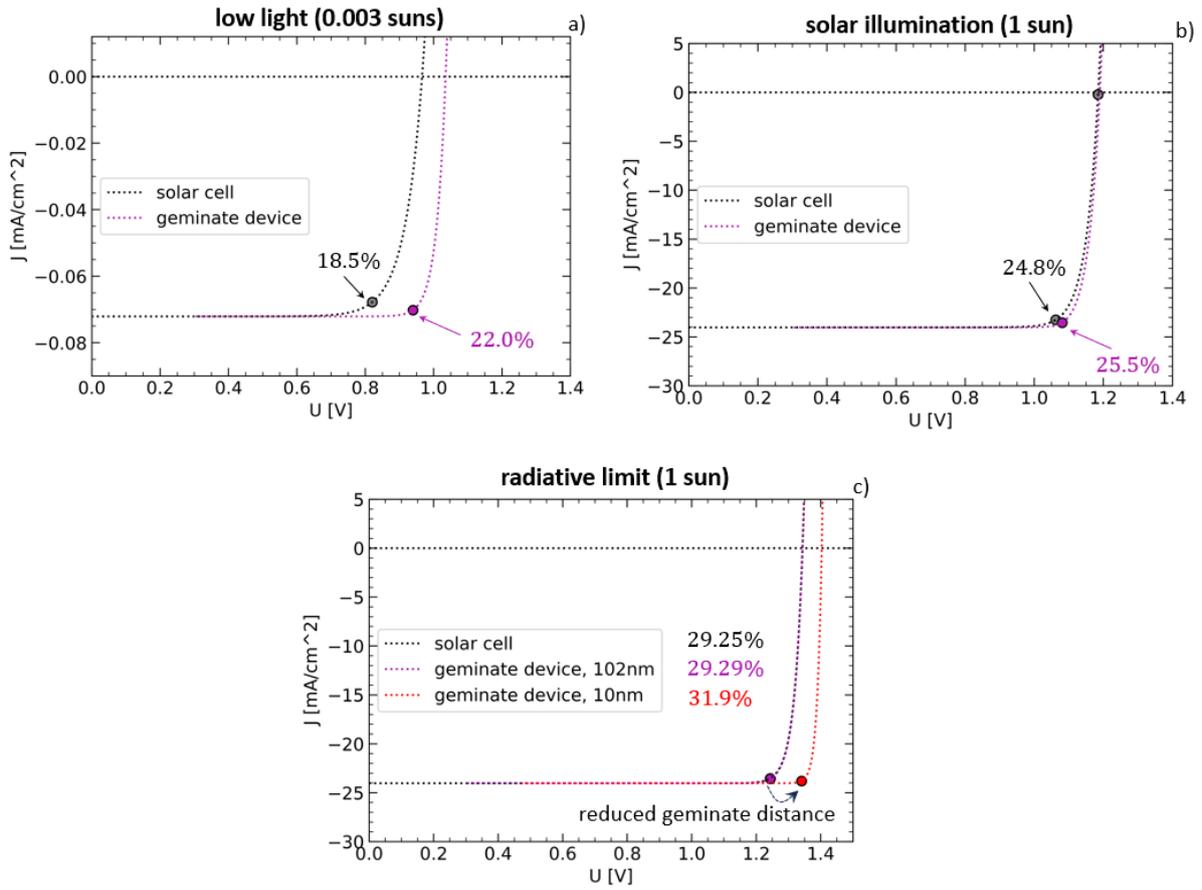

*Figure S5: Current-Voltage Curve from Detailed Balance Analysis and the Efficiency Gain of Geminate Devices. Comparisons between conventional solar and geminate devices for a) illumination with 0.003 suns, b) illumination with 1 sun, and c) illumination with 1 sun and only radiative recombination (radiative limit). The geminate distance is reduced from 102 nm to 10 nm.*

The current-voltage curve can be approximated as Equation S57, which shows that the ideality factor $n_{ID} = h/m$ depends on the domination PL type ($h$) and no the recombination order $m$:

$$J \sim exp\left(\frac{m}{h}\frac{V}{k_B T}\right) \tag{S57}$$